\documentclass[twocolumn,aps,superscriptaddress]{revtex4-2}

\usepackage{amsmath, amssymb, amsthm, graphicx, epsfig, fancyhdr,epsfig,multirow}
\usepackage[utf8]{inputenc}
\usepackage{amsmath}
\usepackage{amsfonts}
\usepackage{amssymb}
\usepackage{xcolor}
\usepackage{comment}
\usepackage{subcaption}
\usepackage[normalem]{ulem}
\usepackage{tabularx}
\usepackage{comment}
\usepackage[left=2cm,right=2cm,top=2cm,bottom=2cm]{geometry}
\usepackage[font=small,labelfont=bf,textfont=it]{caption}
\usepackage {ulem}

\usepackage{color}

\begin{document}
\title{Evolution of topological charge through chiral anomaly transport}
\author{Zilin Yuan}
\email{yuanzilin20@mails.ucas.ac.cn}
 \affiliation{School of Nuclear Science and Technology, University of Chinese Academy of Sciences, Beijing, 100049,
  P.R. China}
\affiliation{Institue of High Energy Physics, Chinese Academy of Sciences,
Beijing, 100049,
  P.R. China}
\author{Anping Huang}
\email{huanganping@ucas.ac.cn}
\affiliation{School of Nuclear Science and Technology, University of Chinese Academy of Sciences, Beijing, 100049,
  P.R. China}
\author{Wen-Hao Zhou}
\email{zhouwenhao@xaau.edu.cn}
\affiliation{Key Laboratory of Nuclear Physics and Ion-beam Application (MOE), Institute of Modern Physics, Fudan University, Shanghai 200433, China}
\affiliation{Shanghai Research Center for Theoretical Nuclear Physics, NSFC and Fudan University, Shanghai $200438$, China}
\affiliation{Faculty of Science, Xi’an Aeronautical Institute, Xi’an 710077, China}
\author{Guo-Liang Ma}
\email{glma@fudan.edu.cn}
\affiliation{Key Laboratory of Nuclear Physics and Ion-beam Application (MOE), Institute of Modern Physics, Fudan University, Shanghai 200433, China}
\affiliation{Shanghai Research Center for Theoretical Nuclear Physics, NSFC and Fudan University, Shanghai $200438$, China}
\author{Mei Huang}
\email{huangmei@ucas.ac.cn}
\affiliation{School of Nuclear Science and Technology, University of Chinese Academy of Sciences, Beijing, 100049,
  P.R. China}

\date{2023.01.13}
\begin{abstract}
Built upon the state-of-the-art model a multiphase transport (AMPT), we develop a new module of chiral anomaly transport (CAT), which can trace the evolution of the initial topological charge of gauge field created through sphaleron transition at finite temperature and external magnetic field in heavy ion collisions. The eventual experimental signals of chiral magnetic effect(CME) can be measured. The CAT explicitly shows the generation and evolution of the charge separation, and the signals of CME through the CAT are quantitatively in agreement with the experimental measurements in Au+Au collision at $\sqrt{s}=200 {\rm GeV}$, and the centrality dependence of the CME fraction follows that of the fireball temperature.
\end{abstract}
\maketitle

{\it Introduction:} The matter-antimatter asymmetry or baryon number asymmetry of our universe (BAU) is tightly related to the the topological $\theta$ vacuum configurations of gauge fields, the electroweak baryongenesis through C and CP-violated sphaleron transition \cite{Kuzmin:1985mm,Shaposhnikov:1987tw,Rubakov:1996vz} is not enough to explain the measurement, which calls for new extension of the additional sources of CP violation.

The $\theta$ vacuum of quantum chromodynamic dynamics (QCD) gauge field characterized by the integer Chern-Simons number \cite{Belavin:1975fg}, the tunneling transition through instanton or sphaleron across the energy barriers leads to non-conservation of the axial current 
\begin{equation}
\partial^\mu j_\mu^5 = 2 \sum_f m_f \langle \bar \psi i \gamma_5 \psi \rangle_A - \frac{N_f g^2}{16\pi^2} F_{\mu\nu}^a \tilde F^{\mu\nu}_a,
\end{equation}
thus the chirality imbalance between the left-handed and right-handed quarks $N_5=N_L-N_R=\int{d^4 x}\partial_{\mu}j^{\mu}_5$ \cite{Witten:1979vv,Veneziano:1979ec,Vicari:2008jw,Schafer:1996wv}.
It has been proposed that local P and CP–odd domains can be formed in heavy-ion collisions or early universe \cite{Kharzeev:1998kz,Kharzeev:1999cz,Kharzeev:2001ev,Kharzeev:2004ey}.
In non-central heavy ion collisions, the strong magnetic field with the largest strength of about $ 10^{14}$ T can be produced, for the system with a net chirality imbalance characterized by $\mu_5$ for massless quarks with charge $Q_e$, there would be a charge current $\bf{J}$ produced along the direction of the magnetic field $\bf{B}$, which is called the Chiral Magnetic Effect (CME) \cite{Fukushima:2008xe}. The CME charge current induces a separation of positively and negatively charged particles perpendicular to the reaction plane. Therefore, the chiral anomaly and CME in the early stage turns into an observable effect of charge separation in the final freeze-out state \cite{Kharzeev:2015znc}. An observation of the CME would verify the fundamental property of QCD, which would provide a natural solution to the baryon number asymmetry of our universe. 


The main observables of CME is the charge azimuthal two-particle correlation:
\begin{equation}\label{mg1}
\begin{split}
\gamma &= \left\langle\cos(\phi_{\alpha} +\phi_{\beta} -2\Psi_{RP} )\right\rangle,  \\
\end{split}
\end{equation}
where $\phi_{\alpha} ,\phi_{\beta} ,\Psi_{RP} $ denote the azimuthal angles of produced charged particles and the reaction plane, respectively, and $\alpha$ and $\beta$ represent either the positive or negative charges. A positive opposite-sign (OS) correlator and a negative same-sign (SS) correlator has been expected to occur for the CME. This feature of CME has been observed by measurements of the correlator $\gamma$ by the STAR Collaboration for Au + Au collisions at $\sqrt{S_{NN}}$ = 200 GeV \cite{STAR:2021mii,STAR:2014uiw} and by the ALICE Collaboration for Pb + Pb collisions at $\sqrt{S_{NN}}$  = 2.76 TeV \cite{ALICE:2012nhw}. 


Difficulty in detecting CME signals is due, on the one hand, to the complexity of the backgrounds, including elliptic flow \cite{Bzdak:2009fc}, resonance decays \cite{Wang:2009kd}, and local charge conservation \cite{Schlichting:2010qia,Wu:2022fwz}. Most of the sign-independent background can be effectively eliminated by calculating the difference between opposite-sign(OS) and same-sign(SS) $\gamma$:
\begin{equation}
    \Delta\gamma=\gamma_{os}-\gamma_{ss}.
\end{equation}
On the other hand, since the CME signal are mostly generated in the early stage of heavy-ion collisions, the final state interactions attenuate the initial CME signal \cite{Ma:2011uma}. Therefore, several hard experimental attempts have been made to detect or even extract the CME signal \cite{Zhao:2019hta,Wang:2018ygc}. For example, recent measurements using the two-plane method \cite{Xu:2017qfs,Chen:2023jhx} have shown that the CME signal contributes no more than 10\% to the total observable \cite{STAR:2021pwb}.


From theoretical simulation side, the CME results from UrQMD and HIJING were inconsistent with the STAR experimental measurement \cite{STAR:2009tro}, and the anomalous-viscous fluid dynamics (AVFD) model \cite{Kharzeev:2022hqz,Shi:2017cpu,Huang:2021bhj} and the AMPT model \cite{Ma:2011uma,Deng:2018dut,Zhao:2022grq} have been developed to interpret the data. 
The AMPT model has an advantage to simulate parton evolution and has been widely used in heavy ion collisions. The anomalous-viscous fluid dynamics (AVFD)\cite{Shi:2017cpu, Huang:2021bhj}, implements the dynamical CME transport in the realistic environment of a relativistically expanding viscous fluid and provides the quantitative link between the CME transport and experimental signal. However, the AVFD is based on hydrodynamics and does not incorporate the signal from the non-equilibrium stage, during which the magnetic field is at its maximum. It has been summarized in \cite{Koch:2016pzl}, that theoretical uncertainties originate from: the initial distribution of axial charges, the evolution of the magnetic field,  the pre-equilibrium dynamics, the hadronic phase and freeze-out conditions. In this work, we provide a theoretical framework called chiral anomaly transport (CAT) to reduce these uncertainties. 
 
\begin{figure}[!h]
\hspace*{-6 mm}
\includegraphics[scale=.45]{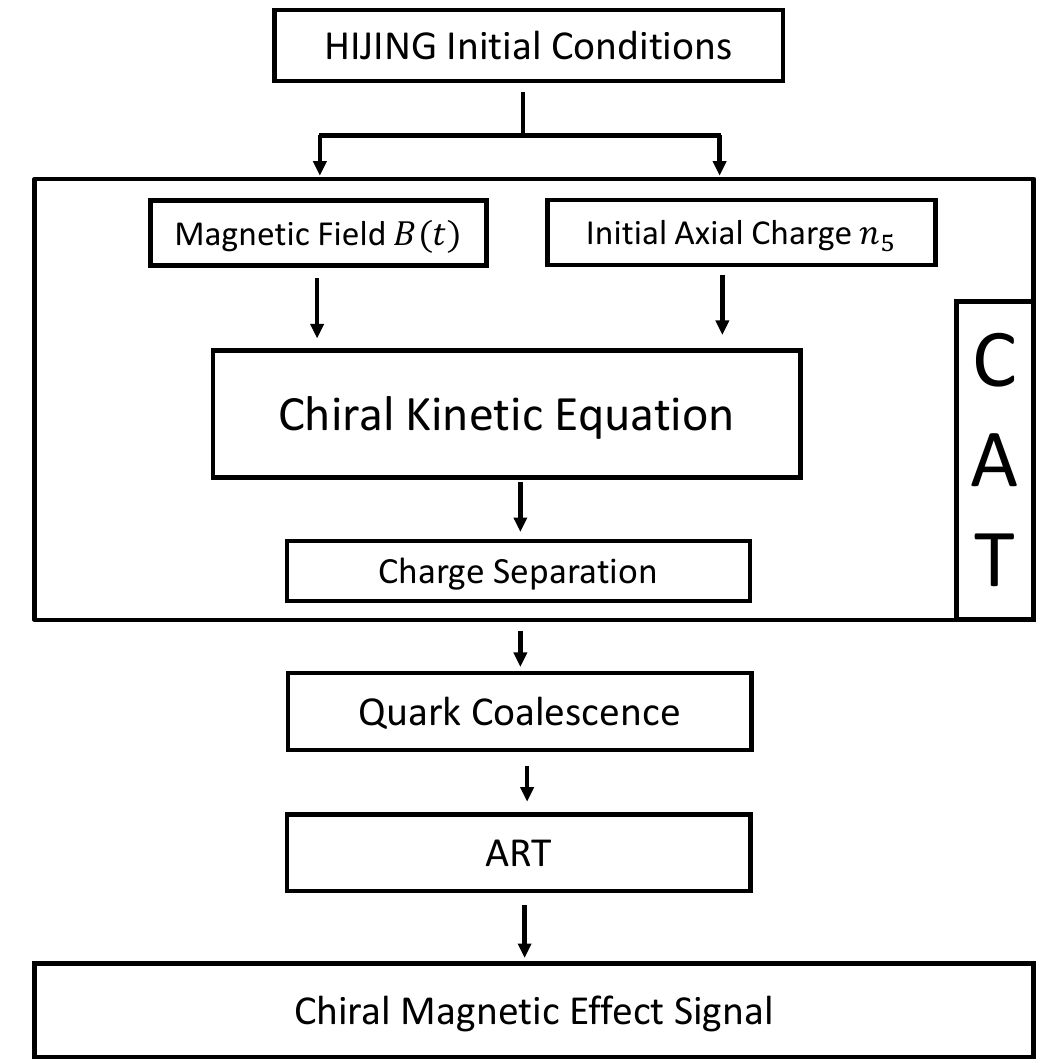}
\caption{(Color online) Illustration of the structure of the chiral anomaly transport model based on AMPT model. The initial nucleons and partons come from HIJING. The main part of CAT corresponding to the parton cascade, which include the magnetic field ,the axial charge and the chiral kinetic theory. All freeze-out partons combine to hadrons and the hadron rescattering is based on the ART model. The CME signal is calculated by the final particle.}
\label{CAT-Structure}
\end{figure}

{\it The Chiral Anomaly Transport (CAT) module:} Built upon the state-of-the-art model AMPT \cite{Lin:2004en}, the CAT module can provide a quantitative connection between the initial topological charge and the eventual experimental signal by simulating the chiral anomalous transport in heavy ion collisions. The structure and flow chart of CAT is illustrated in Fig. \ref{CAT-Structure}.

AMPT consists of four parts: 
The initial distribution of quarks is sampled by the HIJING module from mini-jets and excited strings, subsequently, the quark distribution is imported into the ZPC module to simulate quark evolution, then hadrons are produced from freeze-out quarks in the Quark Coalescence module, finally, the hadrons enter the hadron rescattering stage, which is modeled in the ART module, and this ultimately gives rise to the observable signals. For more details, see ref\cite{Lin:2004en,Lin:2021mdn}. 

 As shown in Fig. \ref{CAT-Structure}, the CAT model which incorporates the chiral kinetic equation, is developed to substitute the ZPC module in the AMPT model. This modification allows for the simulation of the dynamical evolution of chiral quarks in the presence of a strong magnetic field within the expanding fireball. In this model, the program begins with a specified initial condition of quarks distribution and chirality imbalance. The initial quark distribution is generated by the HIJING module. Chiral quarks undergo dynamical evolution within a strong magnetic field, giving rise to a Chiral Magnetic Effect (CME) current. This current, in turn, results in charge separation within the Quark-Gluon Plasma (QGP). Eventually, after undergoing the Quark Coalescence and hadron  rescattering stages, this charge separation manifests as a dipole term within the azimuthal angle distribution of positively and negatively charged hadrons. Such a dipole signal can be measured by the difference between same-sign (SS) and opposite-sign (OS) charged hadron pair correlations $\Delta \gamma = \gamma_{os}-\gamma_{ss}$.

{\it Chiral kinetic equation:} The core of CAT is to solve the chiral kinetic equation, for massless particles in magnetic field it takes the following form\cite{Son:2012wh, Stephanov:2012ki, Huang:2018wdl, Zhou:2018rkh,Sun:2018idn}:
 \begin{eqnarray} \label{ckt}
 \left[ \partial_t + {\dot{\bf{x}}}\cdot   {\nabla}_{x}  +  {\dot{\bf{p}}} \cdot \nabla_{p} \right] f_i({\bf{x}},{\bf{p}},t) = C[f_i]  \,\, ,
 \end{eqnarray}
 \begin{eqnarray} \label{eom}
\sqrt{G}\dot{\mathbf{x}}  =\mathbf{v} +q_{i} \mathbf{B}(\mathbf{v} \cdot \mathbf{b}),\,\, \sqrt{G}\dot{\mathbf{p}}  =q_{i}\mathbf{v}\times \mathbf{B}.
\end{eqnarray}
where $\mathbf{x}$ and $\mathbf{p}$ represent position and momentum vectors respectively, $q_{i}$ is the quark charge, and $\mathbf{b}=h \frac{\mathbf{p}}{2|\mathbf{p}|^{3}}$ denotes the Berry curvature, where $ h = \pm 1$ represents the helicity, and $\mathbf{B}$ stands for the magnetic field. The factor $\sqrt{G}=1+h \frac{\hbar \mathbf{B} \cdot \mathbf{p}}{2|\mathbf{p}|^{3}}$. In the above, $f_i$ represents the distribution function of a specific quark denoted by $i=(\text{flavor}, h)$, where $\text{flavor}$ can take $u, d, s$. In the CAT module, the distribution function $f_i$ is discretized by the formula of $N_i$ point particles (partons or quarks), i.e $f_{i}(\mathbf{x}, \mathbf{p}, t)\approx\sum^{N_i}_{j=1} \delta(\mathbf{x}-\mathbf{x}_{j}(t))\delta(\mathbf{p}-\mathbf{p}_{j}(t))$. In the classical method, the properties of a point particle are expressed by its phase space coordinates $(\mathbf{x}_{j}, \mathbf{p}_{j})$. However, for chiral fermions, the phase space should be extended to include the helicity or chirality parameter $(\mathbf{x}_{j}, \mathbf{p}_{j}, h_{j})$. As done in the CAT module, we will employ the particle simulation method to numerically solve the kinetic equation for modeling the evolution of chiral quarks under a strong magnetic field.


CAT solves the kinetic equation Eq.(\ref{ckt}) step by step from initialization. In each time interval, CAT firstly selects the eligible partons that have formed but have not yet frozen out,
Secondly, CAT updates the position and momentum of partons based on the equations of motion in Eq.(\ref{eom}). These equations are valid only when $|p|^2 \gg eB$. To address this, we set a momentum threshold; when the square of the momentum is smaller than twice the magnetic field strength (i.e. $|p|^2 < 2eB$), the quantum correction terms in Eq.(\ref{eom}) are omitted, and the partons behave as classical particles subjected to the Lorentz force. Thirdly, the momentum of partons is updated once again due to collision effects. In CAT, we model binary elastic collisions with a cross-section of $\sigma=3$ mb. A collision between two partons occurs when their distance is smaller than this cross-section value $d < \sqrt{\sigma/\pi}$, and their momenta are updated after collision. At the end of each time interval, eligible partons will freeze out, while the remaining partons proceed to the next time interval. This loop continues until it reaches the largest freeze-out time. Afterward, all the partons will be sent to the  quark coalescence process.




 In this kinetic transport process, the charged chiral particles tend to move along the magnetic field. The direction of motion is decided by their charge and helicity. So that, we expect the charge separation occur in the case of the chiral anomaly. And for the same charge, the helicity will split in space automatically. But noticing that quarks and their anti-quark with the same helicity have opposite chirality and the opposite charges leading to opposite splitting. The possibility of a localized chemical potential and chirality depends on the quark distribution.

{\it Initial and evolution of magnetic field:} Following Ref \cite{Chen:2021nxs}, the initial electromagnetic field at $t_{0}=0\,\rm{fm}$ is calculated from spectator protons in HIJING as below:

\begin{equation}\label{mg1}
\begin{split}
e\mathbf{E}_{0}(r)=\alpha_{EM}\sum_{n } Z_{n}
\frac{R_{n}(1-\mathbf{v}_{n}^{2})}{(\mathbf{R}_{n}^2-[\mathbf{v}_{n}\times \mathbf{R}_{n}]^{2})^{3/2} },\\
e\mathbf{B}_{0}(r)=\alpha_{EM}\sum_{n } Z_{n}
\frac{\mathbf{v}_{n}\times \mathbf{R}_{n}(1-\mathbf{v}_{n}^{2})}{(\mathbf{R}_{n}^2-[\mathbf{v}_{n}\times \mathbf{R}_{n}]^{2})^{3/2} },
\end{split}
\end{equation}
where $R_{n}=r-{r}'_{n}$ is the relative position vector from a field point r to a source point ${r}'_{n}$ at the initial time $t_{0}$, and $\alpha_{EM}$ is the
EM fine-structure constant, defined as $\alpha_{EM} = {e^2}/{4\pi}\approx 1/137$. It should be noticed that the direction of magnetic field in CAT is opposite to the y-axis, due to the initial condition of target and projectile nucleons.


The evolution of magnetic field in parton cascade part relates to time \cite{McLerran:2013hla,Deng:2012pc}, as following
\begin{equation}
\label{evolution of mg}
    e\mathbf{B}(t,\mathbf{x})=\frac{e\mathbf{B}(0,\mathbf{x})}{1+(\frac{t}{\tau_B})^{2}},
\end{equation}
where $\tau_B = 0.4$ fm is an effective lifetime of magnetic field.

{\it Chirality, helicity and initial axial charge:}
It should be noted that the helicity $h=\pm$ and chirality $\chi=R/L$  are the same for quarks but opposite for anti-quarks. Their densities are denoted as follows:
$n_{R}=n_{+},n_{L}=n_{-},\bar{n}_{R}=\bar{n}_{-},\bar{n}_{L}=\bar{n}_{+}$. The net particles density 
$n=\left \langle \bar{\psi} \gamma^{0}\psi \right \rangle =(n_{R} -\bar{n}_{R})+(n_{L}-\bar{n}_{L})$
could be expressed as $n=(n_{+} - \bar{n}_{-})+(n_{-}-\bar{n}_{+})$. 
And the net chirality 
 $n_{5}=\left \langle \bar{\psi} \gamma^{0}\gamma^{5}\psi \right \rangle =(n_{R} -\bar{n}_{R})-(n_{L}-\bar{n}_{L})$ could be expressed as $n_{5}=(n_{+} + \bar{n}_{+})-(n_{-}+\bar{n}_{-})=\left \langle h \right \rangle$.
For convenience, we just consider the helicity of quark and anti-quark in the program.
At initial stage, $N_{+}=(N_{\text{total}}+N_{5})/2$ particles are randomly selected as positive helicity, and the remaining $N_{-}=(N_{\text{total}}-N_{5})/2$ particles are assigned as negative helicity, where $N_{5}=\int d^{3}x n_{5}$ represents the total chirality.

The initial local chirality density $n_{5} =\mu_{5}^{3}/(3\pi^{2})+(\mu_{5}T^{2})/3$, with $T$ the local temperature and chiral chemical potential $\mu_{5}$, can be either positive or negative with equal chance from event to event. $\mu_{5}$ can be calculated through the sphaleron transition rate $\mu_{5}=\sqrt{3\pi}\sqrt{\left(320N_{f}^{2}\Gamma_{ss}/T^{2}-T^{2}/3\right)}$ \cite{Chao:2013qpa}. At finite temperature and under magnetic field, the sphaleron transition rate has the form of $\Gamma_{ss}(B,T)=\frac{(g_{s}^{2}N_{c})^{2}}{384\sqrt{3}\pi^{5}}\left(eBT^{2}+15.9T^{4}\right)$ in Ref.\cite{Basar:2012gh}, and the chiral chemical potential can be roughly estimated as $\mu_5\sim \left(a T+  b \sqrt{eB}\right)$. It should be emphasized that, $\mu_5$ can be positive or negative in local domains created event-by-event through heavy-ion collisions, one can understand $\mu_5$ in the meaning of average variance $\sqrt{\langle\mu^{2}_{5}\rangle_\text{event}}$. Similarly, the chiral charge number $N_5$, in the CAT module, is equivalent to the average variance of chiral charge number for all of the events defined in \cite{Huang:2021bhj}, specifically defined as $N_5 = \sqrt{\langle N^{2}_{5}\rangle_\text{event}}$, which remains positive and constant for each event.

\begin{figure}[!h]
\hspace*{-6 mm}
\includegraphics[scale=0.35]{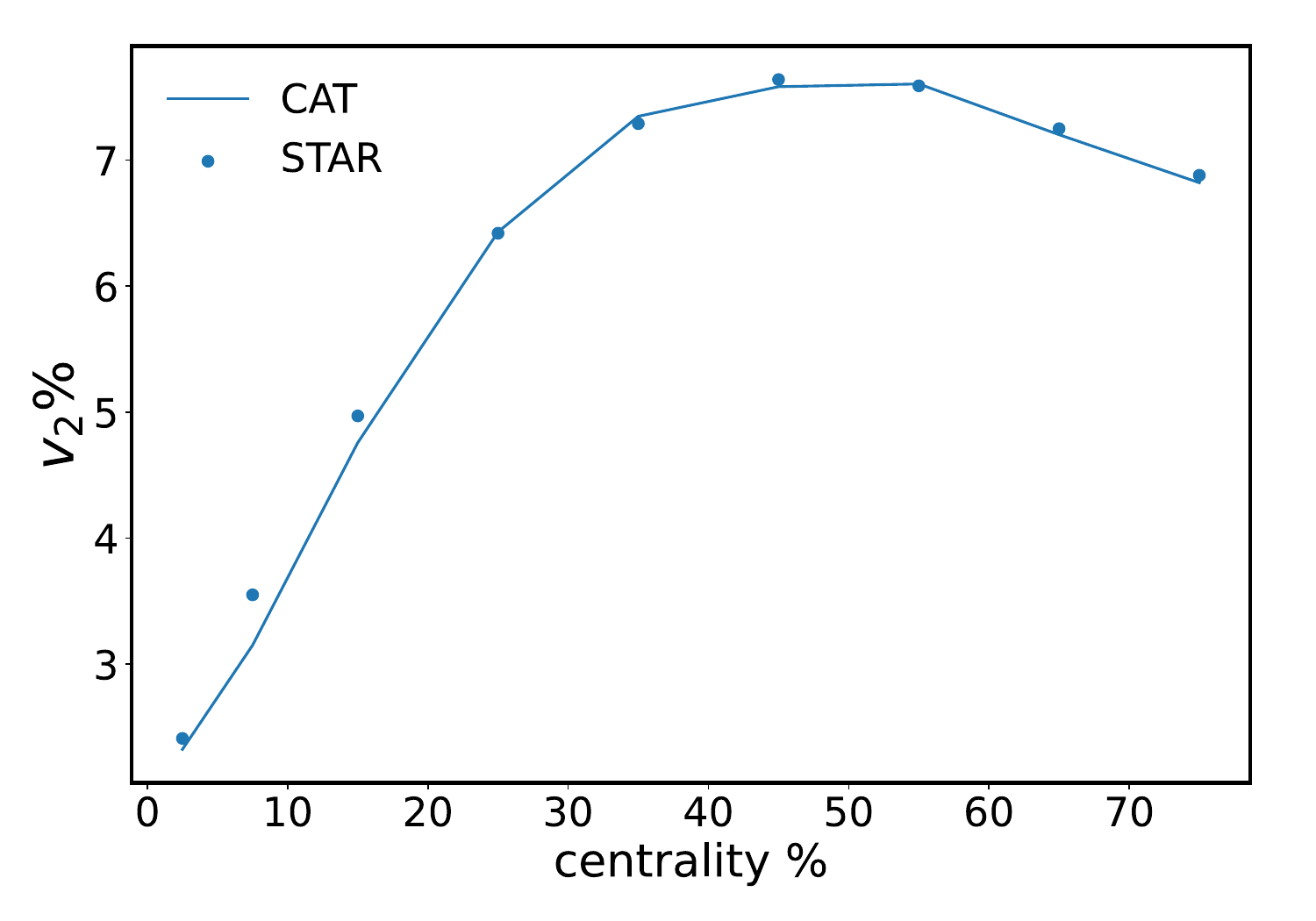}
\caption{(Color online) The CAT result (curve) on the centrality dependence of two-paricle $v_2$ for charged hadrons in Au+Au collisions at 200\,${\rm GeV}$, compared with the STAR data (circles). }
\label{auv2pt}
\end{figure}

\begin{figure}[!h]
\hspace*{-6 mm}
\includegraphics[scale=0.4]{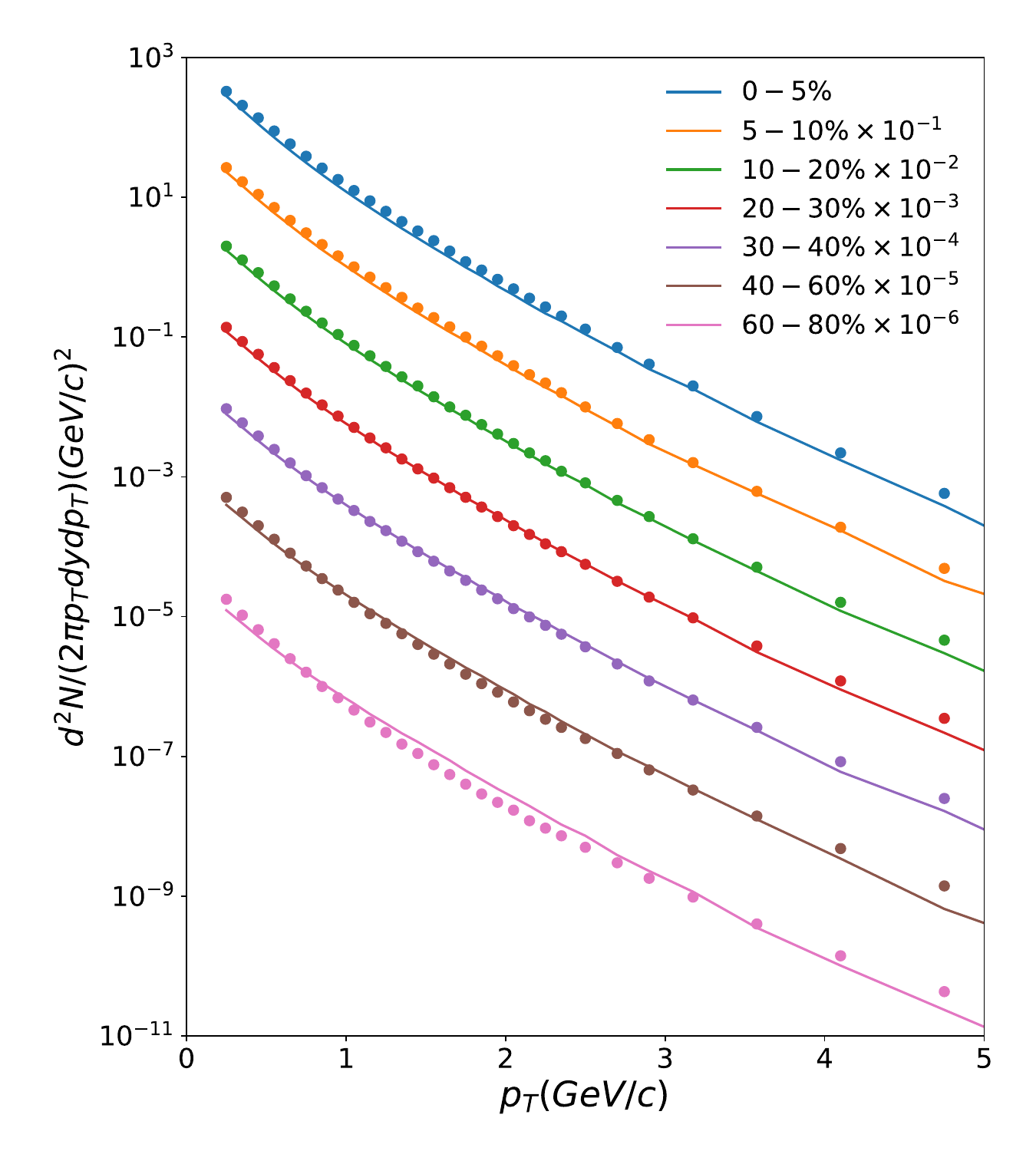}
\caption{(Color online) The CAT results (curves) on $p_T$ distributions of $\frac{1}{2\pi p_T}\frac{d^2 N}{dp_T d\eta }$ for multiplicity of charged hadron of $\left |\eta \right | < 0.5$ for different centralities in Au+Au collisions at $200\,{\rm GeV}$, compared with the STAR data (circles).}
\label{dndptau}
\end{figure}

{\it Results:} The background of the CME is known to be proportional to the ratio of elliptic flow to multiplicity, i.e. $\Delta \gamma \sim - v_{2}/N$. In order to reproduce the reliable background, we first calculate the elliptic flow $v_{2}\{2\}$ of charged particles using the two-particle correlation method \cite{Borghini:2001vi}. Fig.\ref{auv2pt} shows the centrality dependence of the elliptic flow of charged particles within the kinetic cut $0.15<p_T<2\,{\rm GeV/c}$ and $ \left | \eta \right | <1 $, where our definition of centrality is as the same as in experiments by using the reference multiplicity distribution. The CAT result is in good agreement with the STAR data. On the other hand, the $p_T$ distributions of multiplicity are calculated within the pseudorapidity window of $ \left | \eta \right | <0.5 $. Fig.\ref{dndptau} shows the CAT results on the $p_T$ distributions for different centrality bins, in comparison with the STAR data. We find that our results can describe the experimental data well. Therefore, we believe that our CAT model provides a reliable description of the CME background. We also check that if we switch off the magnetic field or $\mu_5$, the elliptic flow and the multiplicity distributions are almost unchanged.

\begin{figure}[!h]
\hspace*{-6 mm}
\includegraphics[scale=0.4]{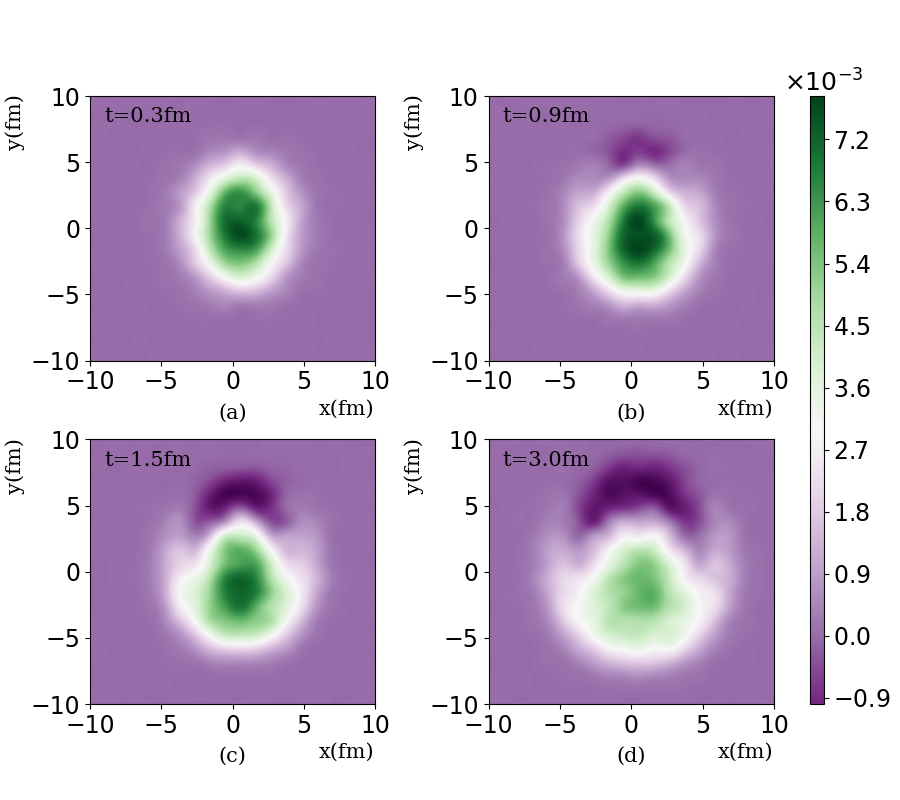}
\caption{(Color online) The CAT results on the distributions of net-charge density in the transverse plane at t= 0.3 fm, 0.9 fm, 1.5 fm, 3.0 fm, respectively, during the partonic evolution at b= 8 fm.}
\label{charge_dis_t}
\end{figure}

The chiral anomaly and the magnetic field can induce a charge current through the chiral magnetic effect. In Fig.\ref{charge_dis_t}, we present the results from the CAT simulation with both magnetic field and the initial net chirality, which shows the time evolution of the net-charge density distribution for the charged quarks within the range $ \left | \eta \right | <1 $ in the transverse plane. This simulation demonstrates the generation of the CME-induced charge separation along the direction of the magnetic field due to chiral anomaly transport. Initially, the net-charge density increases and then slowly decreases with the decay of the magnetic field and the expansion of the QGP. We have verified that this charge separation disappears when the chiral magnetic effect cannot occur, which happens when the chiral chemical potential equals zero.

\begin{figure}[!h]
\hspace*{-6 mm}
\includegraphics[scale=0.35]{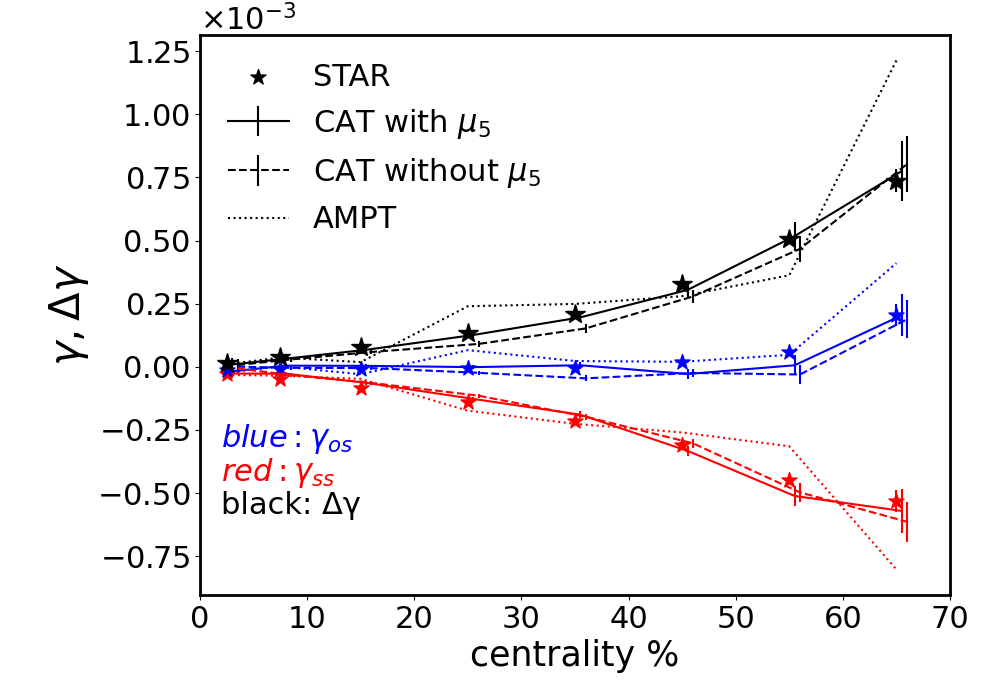}
\caption{(Color online) The CAT results on the centrality dependences of the correlations of opposite-sign $\gamma_{os}$, same-sign $\gamma_{ss}$ and the difference $\Delta \gamma = \gamma_{os}-\gamma_{ss}$ in Au+Au collisions at 200 GeV, compared with the STAR data (stars) and the AMPT result (dot curves), where the different colors represent different correlations. The CAT results are calculated by the CAT with and without chiral chemical potential $\mu_5$, shown by solid and dashed curves, respectively.}
\label{gamma with out miu5}
\end{figure}

\begin{figure}[!h]
\hspace*{-6 mm}
\includegraphics[scale=0.35]{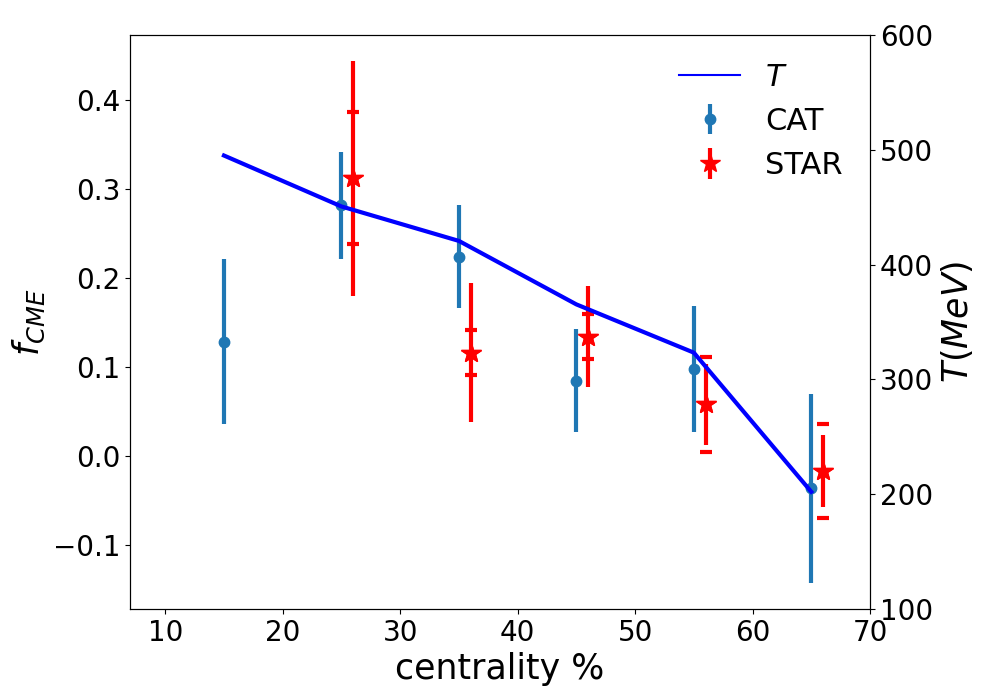}
\caption{The CAT result (circles) on the centrality dependence of the CME fraction $ f_{CME} $, compared with the STAR data (stars) \cite{STAR:2021pwb} and the centrality dependence of the average temperature of fireball(line).   }
\label{fcme1}
\end{figure}

The generation of the chiral magnetic effect contributes to the total two-particle correlation $\gamma_{total}$. On the other hand, the chiral magnetic effect does not occur in the absence of chiral anomaly. Therefore, only the background $\gamma_{bkg}$ can be obtained without the chiral anomaly, which is the total background including the flow background, the non-flow background, and the effect of the Lorentz force. Fig.~\ref{gamma with out miu5} shows the centrality dependences of the opposite-sign correlation $\gamma_{os}$, the same-sign $\gamma_{ss}$ correlation and the correlation difference $\Delta \gamma = \gamma_{os}-\gamma_{ss}$ in Au+Au collisions at 200 GeV, compared with the AMPT model result \cite{Ma:2011uma,Deng:2018dut} and STAR data \cite{STAR:2009tro}. We observe that the CAT model with the chiral anomaly outperforms the other cases in describing the STAR data. In this optimal case, we take 
\begin{equation}
\mu_5 = (2.1T + \sqrt{eB})\,\rm{GeV},
\end{equation} 
and its magnitude is consistent with the effective chiral chemical potential induced by axial-vector interactions \cite{Yu:2014sla,Shao:2022oqw}. The chiral anomaly influences both the same sign and opposite sign correlation, with different significance for each centrality. For instance, in the $20-50\%$ centrality, $\gamma_{ss}$ increases with $n_5$, while $\gamma_{os}$ remains almost unchanged. We notice that the correlations are more sensitive to axial anomaly in middle centrality $20 - 50 \%$, but in centrality $60 - 70 \%$, the difference is negligible within the error. Compared to the previous AMPT study, we observe that the CME leads to nearly zero change of $\gamma_{ss}$ in the 0-50\% centrality and rapidly increase $\gamma_{os}$ in 50-70\%, more consistent with the STAR data. In general, we observe a difference in $10-70\%$ centrality with and without the chiral anomaly, indicating the presence of the CME signal.

In order to extract the contribution of the CME signal to the total $\gamma$ correlation,  the CME signal fractions is defined as,
\begin{equation}
f_{CME}=\frac{\Delta\gamma_{CME}}{\Delta\gamma_{total}}=\frac{\Delta\gamma_{total}-\Delta\gamma_{bkg}}{\Delta\gamma_{total}}.
\end{equation}
Fig.~\ref{fcme1} shows the CAT result on the centrality dependence of the CME fraction $ f_{CME}$, compared with the STAR data \cite{STAR:2021pwb}. The CME fraction decreases from central to peripheral collisions, in good agreement with recent STAR measurement. We observe an increase in CME signal from central to peripheral collisions, but a decrease in CME signal fraction, indicating greater background in peripheral collisions.  We also noticed that the tendency of the CME fraction $f_{CME}$ follows the tendency of centrality dependence of the temperature. 





{\it Conclusion:} Based on the AMPT model, we develop a new module of CAT to solve the chiral kinetic equation in order to trace the evolution of the initial topological charge under magnetic field, and trace the separation of charged particles induced by CME, thus provides a quantitative connection between the initial topological charge and the eventual experimental signal of CME in heavy ion collisions. 
  
For Au+Au collision at energy 200 GeV, the magnetic field is calculated from HIJING module and the initial chirality imbalance is given as $\mu_5 = (2.1 T + \sqrt{eB})\,\rm{GeV}$, then the CAT module solves the evolution of the chial anomaly with the decay of magnetic field. The CAT simulation shows the development of the CME-induced charge separation along the direction of the magnetic field due to chiral anomaly transport, and the results of OS correlator $\gamma_{os}$ and SS correlator $\gamma_{ss}$ as well as their difference  $\Delta\gamma$ are quantitatively in consistent with the STAR measurement.  The background of CME $\gamma_{bkg}$ is regarded as in the case without initial chiral anomaly, and the signal of CME fraction $f_{CME}$ decreases from central to peripheral collisions, which is in good agreement with recent STAR measurement. It is observed that the OS/SS correlator of the CME signal increases from central to peripheral but the signal of CME fraction decreases from central to peripheral, and the tendency of the CME fraction $f_{CME}$ follows the tendency of centrality dependence of the temperature. 

For next step, we will investigate the evolution of chiral anomaly in isobar collisions and Au+Au collision at low energies.

\section*{Acknowledgments}

This work is supported in part by the National Natural Science Foundation of China (NSFC) Grant Nos: 12235016, 12221005, 12147150, 12375121, 12205309 , 12325507, 12147101 and the Strategic Priority Research Program of Chinese Academy of Sciences under Grant No XDB34030000, the start-up funding from University of Chinese Academy of Sciences (UCAS), the Fundamental Research Funds for the Central Universities, and Natural Science Basic Research Program of Shaanxi (Program No. 2023-JC-QN-0267).


\bibliographystyle{unsrt}
\bibliography{reference.bib}

\end{document}